# A Facile Process to Fabricate Phosphorus/Carbon Xerogel Composite as Anode for Sodium Ion Batteries

Changyu Deng[1] and Wei Lu[1,2,*]

[1] Department of Mechanical Engineering, University of Michigan, Ann Arbor, MI 48109, United States

[2] Department of Materials Science & Engineering, University of Michigan, Ann Arbor, MI 48109, United States

*e-mail: weilu@umich.edu

## Abstract

Sodium ion batteries become popular due to their low cost. Among possible anode materials of sodium ion batteries, phosphorus has great potential owing to its high theoretical capacity. Previous research that yielded high capacity and long duration of phosphorus anode used expensive materials such as black phosphorus (BP) and phosphorene. To take advantage of the low cost of sodium ion batteries, we report a simple and low-cost method to fabricate anode: condensing red phosphorus on carbon xerogel. Even with large particle size (~ 50 μm) and high mass loading (2 mg cm$^{-2}$), the composite cycled at 100 mA g$^{-1}$ yielded a capacity of 357 mA g$^{-1}$ or 2498 mAh g$_P^{-1}$ based on phosphorus after subtracting the contribution of carbon. The average coulombic efficiency is as high as 99.4%. When cycled at 200 mA g$^{-1}$, it yielded a capacity of 242 mAh g$^{-1}$ or 1723 mAh g$_P^{-1}$, with average degradation rate only 0.06% in 80 cycles. Our research provided an innovative approach to synthesize anodes for sodium ion batteries at extremely low cost, with performance exceeding or comparable to state-of-the-art materials, which will promote their commercialization.

# 1. Introduction

Sodium ion batteries (SIBs) have been attracting much attention due to abundant resources of precursors and the potential low cost. Phosphorus has a great potential to serve as the anode of sodium ion batteries owing to its high theoretical capacity (2596 mAh g$^{-1}$ by forming Na$_3$P)[1]. However, the large radius of sodium ion causes a large volume change during intercalation/deintercalation and material degradation, which makes it difficult to achieve and maintain high capacity after cycling.

**Table 1.** A comparison of some state-of-the-art phosphorus anode materials. Metrics include (from left to right) electrode mass loading, current density, reversible specific capacity calculated based on P, degradation rate, average coulombic efficiency (CE) and cost. Bold numbers denote the best performance. Capacity and degradation depend on the current density, so we marked several top performers. The dash denotes no reported value.

| Ref. | Materials | Mass loading (mg cm$^{-2}$) | Current density (A g$_P^{-1}$) | Specific capacity (mAh g$_P^{-1}$) | Degradation (% cycle$^{-1}$) | CE (%) | Cost |
|---|---|---|---|---|---|---|---|
| [2] | Graphene, phosphorene | 1.15 | 0.05 | 2440 | 0.16 | 96[a] | $$$ |
|  |  |  | 8 | 1450 | 0.16 | 97.6 |  |
| [3] | Graphene aerogel, red P | - | 0.26 | 2085 | 0.11 | 95.9 | $$$ |
|  |  |  | 2.6 | 1638[a] | 0.17 | 93.3 |  |
| [4] | Mesoporous carbon CMK-3, red P | 1~1.3 | 0.52 | **2549** | 0.26[a] | - | $$$ |
|  |  | **2.2** |  | **2559**[a] | 0.46[a] |  |  |
| [5] | CNTs, red P | - | 0.143 | 1675 | 2.34 | - | $$ |
| [6] | Multiwalled CNTs, ketjenblack, black P | 1.5 | 0.416 | 2011 | 0.20 | >**99**[a] | $$$ |
| [7] | Graphene, black P | 2 | 1 | 1504 | **0.03**[b]~0.08[c] | >**99**[a] | $$$ |
| This work | Carbon xerogel, red P | 2 | 0.92 | **2498** | 0.10 | **99.4** | $ |
|  |  |  | 1.84 | **1723** | **0.06**~0.18 | 99.2 |  |

Note: a) estimated from plots
b) calculated by the 1$^{st}$ and 500$^{th}$ cycles
c) reported by the authors for the first 100 cycles.

Various methods have been attempted to increase the capacity of phosphorous anode [8]. For instance, Sun et al. [2] mixed graphene and phosphorene to yield a specific capacity of up to 2440 mAh g$_P^{-1}$ (calculated based on phosphorus) and a degradation rate of 0.16% per cycle, at current density of 50 mA g$_P^{-1}$ and total mass loading of 1.15 mg cm$^{-2}$. Gao et al. [3] integrated red phosphorus with graphene aerogel to yield a capacity of 2085 mAh g$_P^{-1}$ at current density of 260 mA g$_P^{-1}$ and a degradation rate of 0.11% per cycle. The initial capacity was 1638 mA g$_P^{-1}$ with a degradation rate of 0.17% per cycle when the current density was increased to 2.6 A g$_P^{-1}$. It is also worthy to note that their average coulombic efficiencies are only 97.3% and 93.3% at the two cycling current densities, respectively. Li et al. [5] mixed red phosphorus (red P) with carbon nanotubes (CNTs) to deliver a reversible capacity of 1675 mAh g$_P^{-1}$ at current density of 143 mA g$_P^{-1}$, with capacity retention of 76.6% over 10 cycles (2.34% degradation per cycle). Besides, black phosphorus (BP) which has higher conductivity than red P is used as anode materials as well [6,7,9]. As listed in Table 1, among these recent discoveries and achievements, anode for sodium ion batteries yield relatively low capacity and fast degradation rate. What is worse, the anodes are composed of extremely expensive materials, such as graphene, BP and phosphorene. These materials have already lost the low-cost advantage of SIBs, impeding the scale-up production and commercialization of SIBs. There have been some attempts on carbon-based anode materials for SIBs [10,11]. Carbon black and hard carbon are commonly used energy storage materials, and they are reported to yield capacity of 200 mAh g$^{-1}$ [12] and 300 mAh g$^{-1}$ [13], yet at a very low current density impractical at all for real applications. To achieve a capacity over 200 mAh g$^{-1}$, nano materials have been used, such as hollow nanowires [14], graphene[15], and hard carbon nanoparticles [16]. To facilitate industrialization, low-cost methods have been explored yet the performance is not ideal; for instance, Shen et al. [17] carbonized wood to serve as anode, and reported 80 mAh g$^{-1}$ at 0.5 C-rate. We aim to improve the capacity more by phosphorus while maintaining low cost.

The principal strategy for making anode with P is to fabricate a conductive base that can help to transfer electrons and sodium ions into or out of phosphorus. This motivated us to use carbon xerogel to build a carbon network. Carbon xerogel is a carbon skeleton with interconnected porous structure and high surface area. Xerogel is similar to and sometimes regarded as a subcategory of aerogel which is attractive as electrodes for electrochemical double layer capacitors and electrosorptive processes [18]. The major difference between xerogel and aerogel is that xerogel is dried at ambient atmosphere, while aerogel is dried by other advanced and costly methods such as supercritical carbon dioxide [19]. Carbon xerogel from polymer pyrolysis can be manufactured at a very low cost. For instance, Li et al. [20] reported a polymer xerogel manufacturing process by simply dissolving and precipitating polyvinyl chloride (PVC) which allows reusing plastic waste.

In this paper, we report a method to condense red P on carbon xerogel to synthesize anode with low cost, high capacity, and good capacity retention. We mixed resorcinol with formaldehyde and carbonized the polymer at 1000 °C to obtain carbon xerogel. Carbon xerogel was sealed with red P in vacuum and heated up to 900 °C to condense phosphorus vapor on the carbon skeleton. The composite took advantage of the high conductivity of carbon and the high theoretical capacity of phosphorus. We achieved a maximum capacity of 2498 mAh $g_P^{-1}$ and degradation rate as low as 0.06% per cycle. In addition, we would like to note that xerogel has higher density than aerogel, which gives a good volumetric capacity as well. The fabricated electrode has a density similar to a graphite electrode. Our method is easy to implement with low cost, ready for commercialization to meet the large demands of energy storage.

## 2. Experimental

2.1 Carbon xerogel

Carbon xerogel (CX) was prepared by a typical procedure [21]. Resorcinol and formaldehyde were mixed with a mole ratio of 1:2. Sodium carbonate was added as catalyst and the mole ratio of resorcinol to sodium carbonate is 50:1. Deionized water was used to dilute the solution such that the mass of resorcinol was 5% of the whole solution. The initial pH of the solution was set as 6.0 by dilute HNO$_3$. The solution was sealed and stirred magnetically for 30 min and kept at room temperature for three weeks (no stirring). After gelation, the gel was immersed in ethanol for three days with fresh solvent replaced daily. The wet gel (Figure 1, left column) was dried and carbonized in a chemical vapor deposition (CVD) furnace with flowing argon gas. We used argon instead of nitrogen to get rid of any influence of nitrogen on the capacity [11]. At a heating rate of 0.5 °C min$^{-1}$, the gel was heated to 65 °C and held for 5 h; then it was heated to 110 °C and held for 5 h. Afterwards, it was heated to 1000 °C at a heating rate of 5 °C min$^{-1}$ and held for 4 h. The flow rate of argon was set as constant 200 SCCM. The product is carbon xerogel (CX), as shown in the middle column of Figure 1.

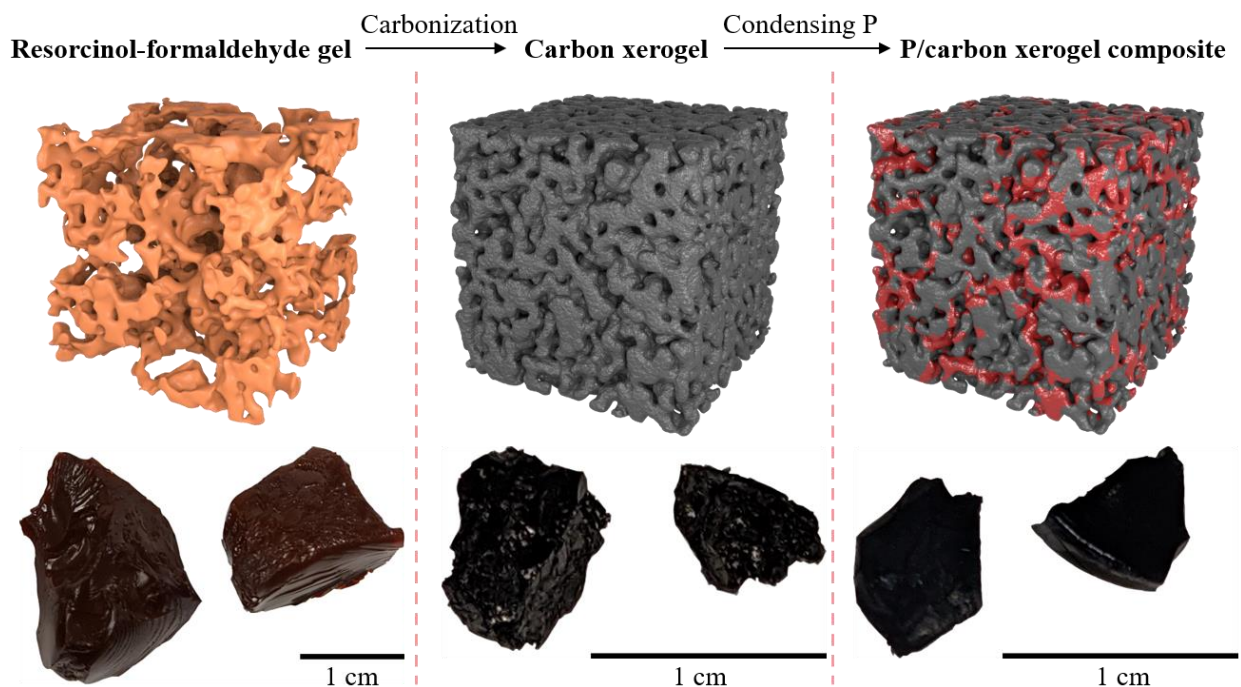

**Figure 1.** Intermediate and synthesized materials. Top row: workflow showing the synthesis of resorcinol-formaldehyde (RF) gel to carbon xerogel (CX) and further to phosphorus/carbon xerogel composite (P@CX). Middle row: illustration of the microstructures. Bottom row: images of the materials. The volume shrinks after carbonization, which means that the porosity reduces. CX and P@CX exhibit similar appearance.

2.2 Phosphorus condensation

0.05 g of CX prepared in the previous step was sealed in a quartz tube with 0.05 g commercial red P. The tube was heated to 900 °C at a heating rate of 4 °C min$^{-1}$, and held for 4 hours for phosphorus to diffuse into CX. Next, the tube was cooled down to 280 °C at a rate of 1 °C min$^{-1}$ and held for 24 h to convert white P that might be generated from red P during the previous high temperature back to red P. After cooled down, the product of phosphorus/carbon xerogel (P@CX) was washed by CS$_2$ to remove any white P formed during condensation (Figure 1, right column).

2.3 Thermogravimetric Analysis

Thermogravimetric analysis (TGA) was conducted by a thermal analysis device TGA 5500 (TA Instruments) to measure the mass ratio of phosphorus in the P@CX composite. Samples were immersed in 25 mL min$^{-1}$ nitrogen gas flow and purged for 30 min at room temperature prior to heating. Assuming that all phosphorus would evaporate when the temperature reached around 400 °C, we heated the samples till 500 °C which was sufficient to cover the range of interest [22]. The heating rate was 2 °C min$^{-1}$.

2.4 Electrochemical characterization

The P@CX composite was granulated into powder by zirconium oxide beads in SpeedMixer. The powder (80 wt.%) was mixed with CMC-Na binder (10 wt.%), super P (10 wt.%) and water to make a homogeneous slurry. The slurry was pasted on 9 μm-thick Cu foil (MTI Corp.) and vacuum-dried for 4 hours at 60 °C. The electrode sheet, which had a total mass loading of about 2 mg cm$^{-2}$, was cut into disks. Next, the electrode disks were assembled to sealed 2032 type coin cells (MTI Corp.) with sodium metal counter and reference electrode and with a separator (Celgard 2320). The assembly was performed in an argon-filled glovebox (MBraun) containing less than 0.1 ppm oxygen and moisture. The electrolyte solution was 1 M sodium perchlorate (NaClO$_4$) dissolved in a mixture (1:1, v/v) of ethylene carbonate (EC, Sigma Aldrich) and dimethyl carbonate (DMC, Sigma Aldrich). Fluoroethylene carbonate (FEC, 5 wt.%) and vinylene carbonate (VC, 1 wt. %) were added to stabilize and reduce the thickness of solid electrolyte interphase (SEI) layers [9,23]. All cells were cycled using a constant current (CC) scheme in a MACCOR battery test system with a 5 min rest between charging and discharging steps.

## 3. Results

The TGA result in Figure 2a shows that the mass of carbon xerogel almost remained constant when the temperature increased up to 500 °C. The constant mass indicates that the gel was sufficiently carbonized before testing, so that they could withstand such a temperature. The mass of the P@CX composite increased gradually initially and then dropped sharply to 87.96% when the temperature reached around 450 °C. The gradual increase was attributed to the explosive sublimation with recoil effect [22,24]. Considering the mass reduction of the carbon skeleton, the mass ratio of phosphorus in the composite is estimated to be 1-87.96%/(1-1.34%)=10.85%.

The pores of the samples are characterized by nitrogen adsorption and desorption. Condensing phosphorus greatly reduces the Brunauer–Emmett–Teller (BET) specific surface area from 591 m$^2$ g$^{-1}$ (CX) to 118 m$^2$ g$^{-1}$ (P@CX). As shown in Figure 2b, phosphorus occupies micropores, especially those below 5 nm; this result accords with another phosphorus condensing measurement [8].

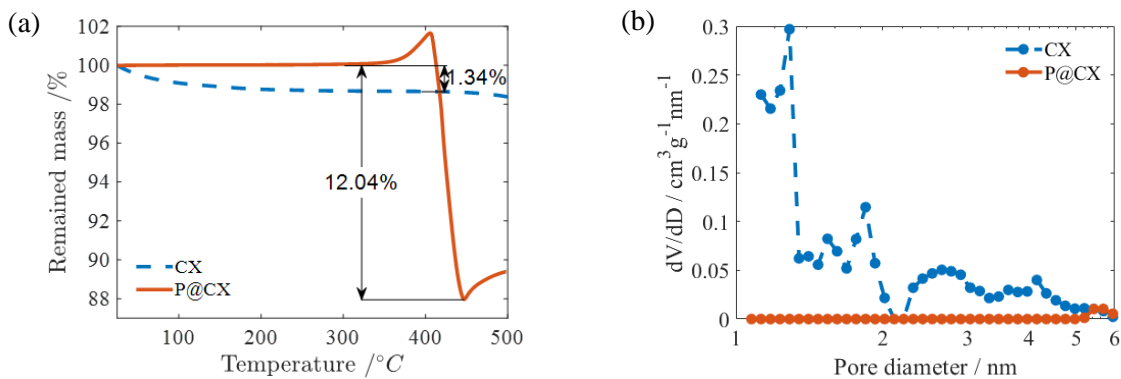

**Figure 2.** Comparison between CX and P@CX by TGA and BET. (a) TGA results. The arrows show that 1.34% of mass is evaporated in CX, while 12.04% of mass is evaporated in P@CX. Here the stationary points are used to calculate the mass loss. (b) Pore size distribution, investigated by the density functional theory (DFT) method. The BET specific surface area of CX and P@CX is 591 $m^2$ $g^{-1}$ and 118 $m^2$ $g^{-1}$, respectively.

The scanning electron microscope (SEM) images of P@CX are presented in Figure 3. From a and b, we can observe that the particle sizes are fairly large: some particles are about 50 μm in diameter. Figure 3c shows the morphology of the fabricated composite electrode, and we can see a porous microstructure. Figure 4 shows that phosphorus is fairly uniformly distributed.

Our strategy is to use the carbon xerogel with interconnected skeleton for enhanced electron conduction and sodium ion transport, rather than relying on making particle size smaller to enhance the transport. Therefore, we purposely did not seek to granulate the material to finer powder. In fact, the large particle size and the high mass loading of 2 mg $cm^{-2}$ place our electrode at a challenging situation. Despite these, we show that the electrode gives excellent performance.

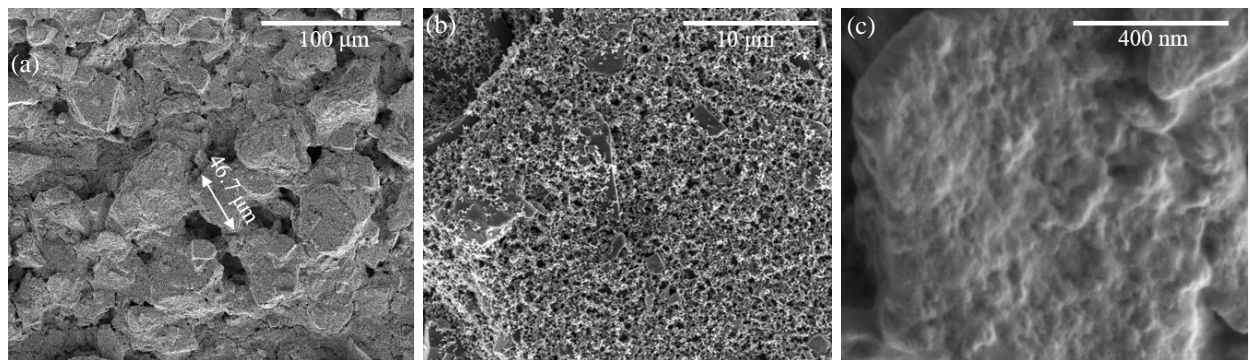

**Figure 3.** SEM images of P@CX. (a) P@CX electrode (containing added carbon black and binder) at low magnification. (b) P@CX electrode at middle magnification. (c) P@CX composite (without carbon or binder) at high magnification.

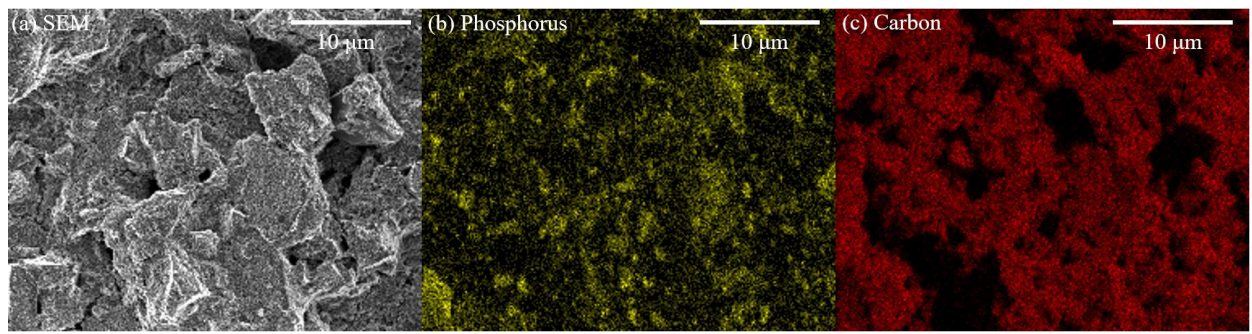

**Figure 4**. Energy Dispersive X-Ray Spectroscopy (EDS) images of P@CX electrode showing P and C distribution.

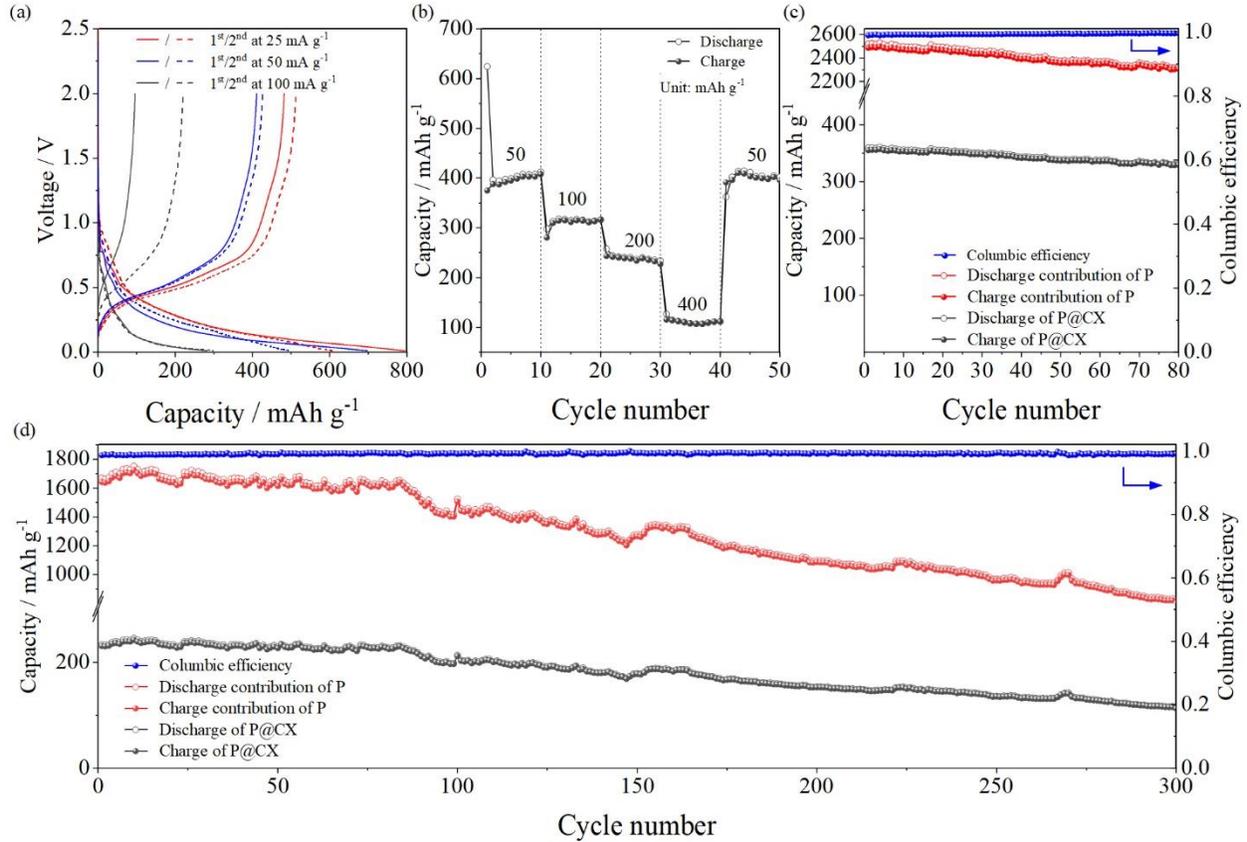

**Figure 5.** Cycling performance of P@CX cells. Current density and capacity are calculated based on the total mass of the composite. (a) Voltage vs. capacity for the first two cycles between 0.01~2 V. (b) Reversible capacity of P@CX cycled at different current densities (between 0.01~1.5 V). (c) Capacity and coulombic efficiency over 80 cycles, cycled at 100 mA g$^{-1}$ (0.92 A g$_P^{-1}$ calculated based on phosphorus) between 0.01~1.5 V. Contribution of P is estimated by Eq. (1). Activation cycles are not included. (d) Capacity and coulombic efficiency over 300 cycles, cycled at 200 mA g$^{-1}$ (1.84 A g$_P^{-1}$ calculated based on phosphorus) between 0.01~1.5 V. Contribution of P is estimated by Eq. (1). Activation cycles are not included.

Figure 5 shows the measured capacity of our P@CX electrode. Discharge refers to sodium intercalation into the P@CX electrode, while charge refers to sodium deintercalation. In Figure 5a, we cycled cells at various current densities between 0.01~2 V. The reversible capacity of the first cycle is lower than that of the second cycle. This implies that the materials needed initial cycling to get activated. Nevertheless, the initial columbic efficiency reaches 60.7% (RC = 482 mAh g$^{-1}$, IC = 794 mAh g$^{-1}$), 58.9% (RC = 411 mAh g$^{-1}$, IC = 697 mAh g$^{-1}$) and 34.1% (RC = 96 mAh g$^{-1}$, IC = 282 mAh g$^{-1}$) at 25 mAg$^{-1}$, 50 mAg$^{-1}$ and 100 mAg$^{-1}$, respectively, much higher than that of CX (less than 25%). Here RC denotes the first cycle reversible capacity and IC denotes the first cycle irreversible capacity. The initial columbic efficiency is calculated by RC/IC. We note that the cell capacity is mainly contributed by the region below 1.5 V. Therefore, we cycled the cells between 0.01~1.5 V in the following plots. In Figure 5b, the cell was cycled at various current densities from 50 mA g$^{-1}$ to 400 mA g$^{-1}$ (based on the mass of P@CX composite), corresponding to 0.46 A g$_P^{-1}$ to 3.7 A g$_P^{-1}$ calculated based on P. The reversible capacity drops from a maximum of 407 mA g$^{-1}$ (based on the composite mass) to a minimum of 107 mA g$^{-1}$ as the current increases, and bounces back to almost the same level of capacity when current returns to a lower magnitude. This shows that the composite is stable at high current. In Figure 5c, we cycled the cell at 25 mA g$^{-1}$ for 2 activation cycles (not shown in the figure) and then at 100 mA g$^{-1}$, or at 0.92 A g$_P^{-1}$ calculated based on phosphorus. Despite the large particle size (~ 50 μm) and high mass loading (2 mg cm$^{-2}$), the P@CX composite yielded 357 mAh g$^{-1}$ maximum reversible capacity. The capacity at the 80th cycle is 330 mAh g$^{-1}$, or about 92.3% of the maximum capacity. Since the capacity fluctuates, it would be inaccurate to calculate the degradation rate using only two points. We fit the capacity vs. cycle number by linear regression, and obtain an average degradation rate of 0.10% per cycle. The average coulombic efficiency is as high as 99.4%. In Figure 5d, we cycled the cell in the same way except that the current density was increased to 200 mA g$^{-1}$, or 1.84 A g$_P^{-1}$ calculated based on phosphorus. The maximum capacity becomes 242 mAh g$^{-1}$. The capacity curve of the cell appears piecewise linear: the cell degrades slowly at the beginning and degrades faster afterwards. The capacity at the 80th and the 300th cycle is 227 mAh g$^{-1}$ and 113 mAh g$^{-1}$, respectively, corresponding to 94.1% and 46.6% of the maximum capacity. The average degradation rate is 0.06% per cycle for the first 80 cycles, and 0.18% per cycle for all 300 cycles. The average coulombic efficiency is 99.2%.

As shown by our TGA result and previous studies on phosphorus/carbon hybrid [22,25], the composite is a physical mixture of P and carbon, thus the capacity is the sum of the two elements. To accurately calculate the phosphorus contribution in the composite, we need to subtract the contribution from carbon (CX). We obtain the capacity of phosphorus by

$$c_P = \frac{c_{P@CX} - c_{CX} m_{CX}}{m_P} = \frac{c_{P@CX} - c_{CX}(1 - m_P)}{m_P} \quad (1)$$

where $c_P$, $c_{P@CX}$ and $c_{CX}$ denote the capacity of phosphorus, P@CX and CX, respectively, calculated based on their own mass. $m_{CX}$ and $m_P$ denote the mass ratio of CX and P in the P@CX composite. To measure the contribution of carbon, we need to separately cycle CX at a similar C-rate as it experiences in the P@CX composite. The maximum capacity of P@CX at 100 mA·g$^{-1}$ is 357 mAh g$^{-1}$, which means that the practical C-rate is 0.28. By charging and discharging CX samples using various current densities to find a C-rate that is close to 0.28, we obtained a capacity of $c_{CX}$=96 mAh g$^{-1}$ (see Figure S1). With Eq. (1), we calculate the maximum capacity of phosphorus to be $c_P$=2498 mAh g$_P^{-1}$ or about 96% of the theoretical capacity. Similarly, for 200 mA·g$^{-1}$, we obtained $c_{CX}$=61 mAh g$^{-1}$ and calculate phosphorus capacity $c_P$=1723 mAh g$_P^{-1}$ or about 66% of the theoretical capacity. Capacity of other cycles can be also estimated by assuming constant contribution ratio of phosphorus to carbon, as shown in Figure 5c and d.

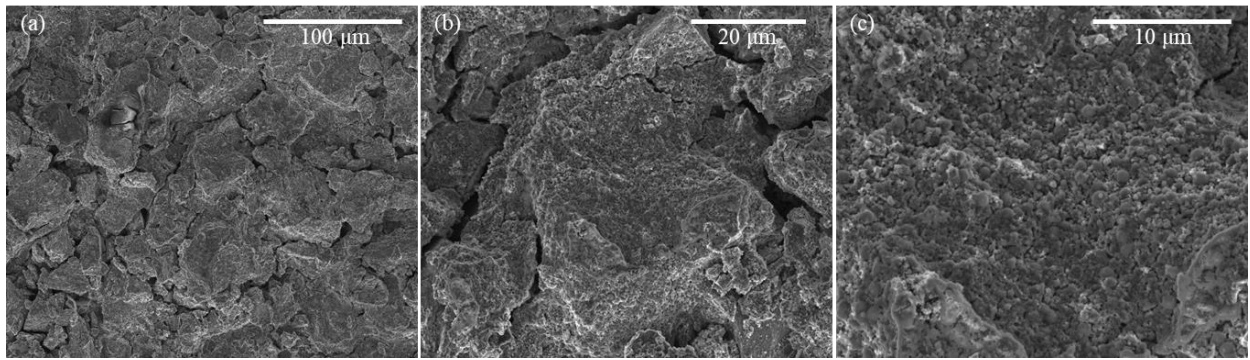

**Figure 6**. SEM images of P@CX electrode (containing added carbon black and binder) after 300 cycles. It is the same sample as in Figure 5d.

We took SEM images of a P@CX sample after 300 cycles (whose capacity is shown in Figure 5d). As shown in Figure 6, there are some minor cracks in the particles, probably since the radius is too large. On the whole, the particles maintain complete shapes, indicating a good buffer performance of carbon xerogel.

## 4. Conclusions

Sodium ion batteries are attractive for the low material cost and the great potential to meet the need for cheap and efficient energy storage systems. In many previous reports of anode materials, expensive materials and sophisticated synthesis procedures are required to achieve reasonable capacity and duration. Otherwise, the capacity, duration and coulombic efficiency are impractical for real applications. We proposed a method to synthesize anode with good performance comparable with state-of-the-art materials but with extremely low cost and simple manufacturing process.

In our P@CX material, the mass ratio of P was measured to be 10.85% by TGA. The reversible capacity is around 357 mAh $g^{-1}$ at 100 mA $g^{-1}$ and 242 mAh $g^{-1}$ at 200 mA $g^{-1}$, with coulombic efficiency as high as 99.4% and 99.2% respectively. Subtracting the contribution of carbon, the capacity of P was estimated to be 2498 mAh $g_P^{-1}$ and 1723 mAh $g_P^{-1}$ at current densities of 0.92 A $g_P^{-1}$ and 1.84 A $g_P^{-1}$, respectively. From the results we can conclude that carbon xerogel could help phosphorus to achieve high capacity.

There still remains much room for improvement. Despite high capacity contribution from P, the capacity of the composite P@CX is not high enough due to low phosphorus ratio. Two major strategies can be implemented in future works: optimizing the carbon skeleton and enhancing the phosphorus condensation conditions. For the carbon xerogel, many parameters (e.g., precursor ratios, temperature and time) can be tuned to achieve higher surface area and thus more adsorption sites. The surface area could also be increased by carbon oxide activation [21]. For the condensation conditions, higher pressure and temperature could potentially increase phosphorus condensation. Our results can help push forward the realization of commercial high capacity anode for sodium ion batteries.

**Acknowledgement**

This research was supported by SAMSUNG Global Research Outreach (GRO) program. We gratefully acknowledge the support. We thank Yixuan Chen, Ruiming Lu and Professor Ferdinand Poudeu for their help with quartz tube sealing. We thank Yuxuan Luan for suggestions on our figures. We thank the Michigan Center for Materials Characterization, the University of Michigan College of Engineering, and NSF grant #DMR-0320740, for funding of the microscope used in this work.

# Supplemental Information

# A Facile Process to Make Phosphorus-doped Carbon Xerogel as Anode for Sodium Ion Batteries

**Table of contents**

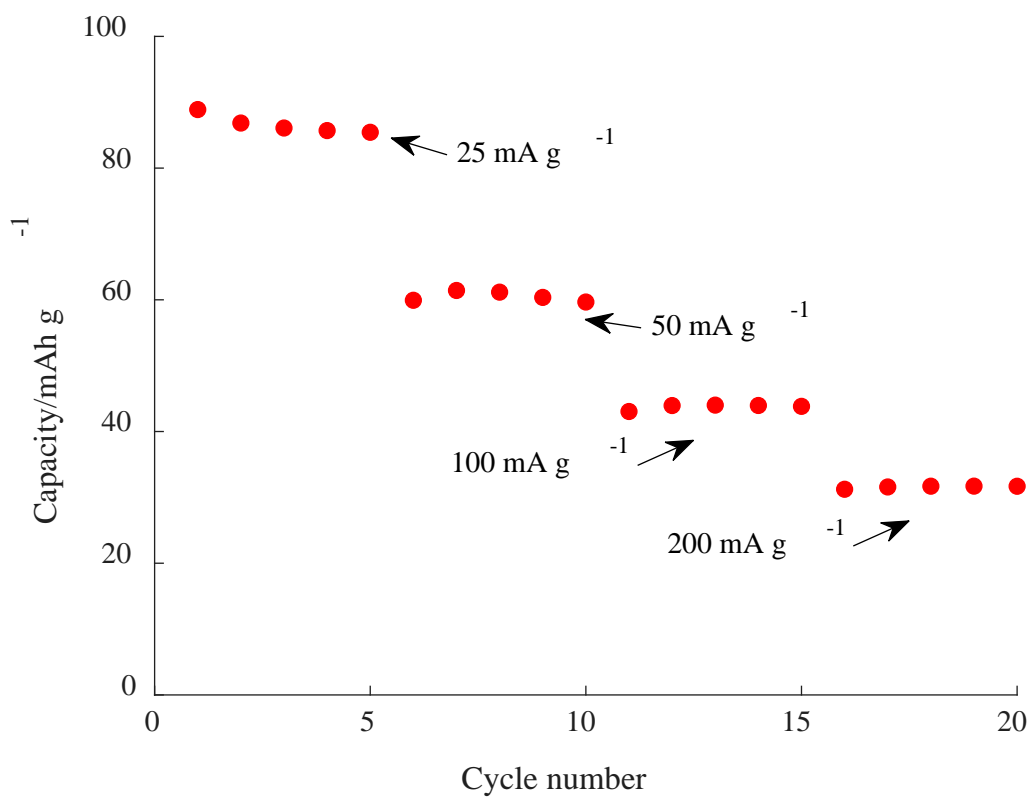

**Figure S1.** Capacity of carbon xerogel (CX). Cycled at 25 mA g$^{-1}$ ~200 mA g$^{-1}$ (calculated based on CX) between 0.01~1.5 V. The maximum capacity values are 96 mAh g$^{-1}$, 61 mAh g$^{-1}$, 44 mAh g$^{-1}$ and 32 mAh g$^{-1}$, respectively, corresponding to C-rates of 0.26, 1.2, 2.3 and 6.2.

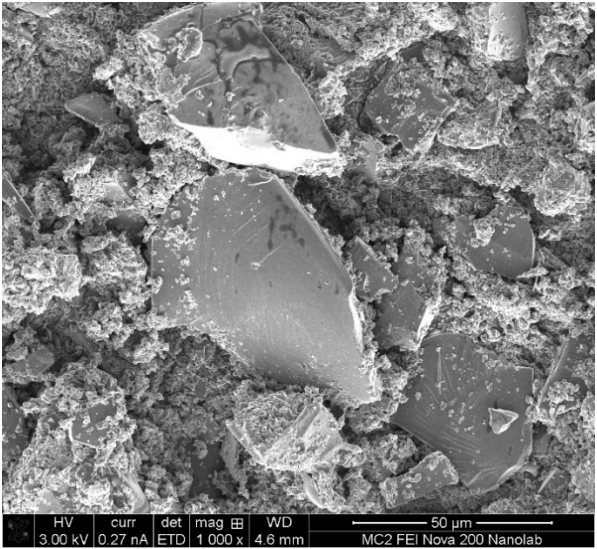
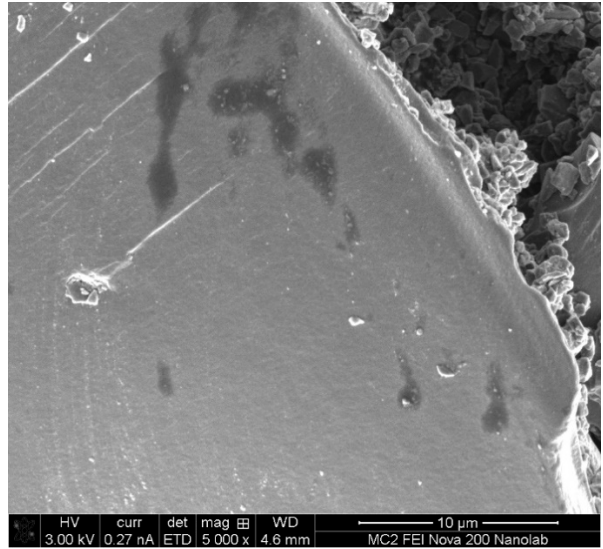
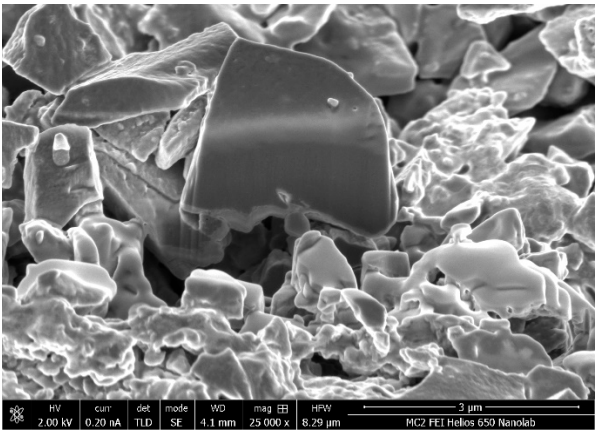
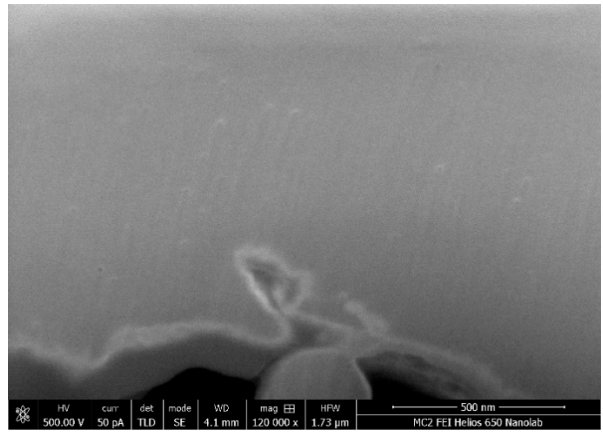
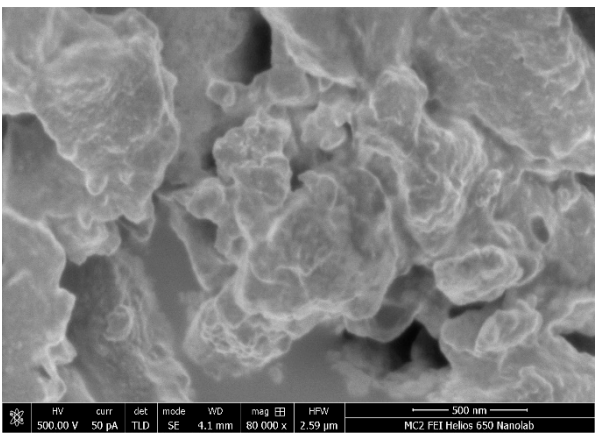
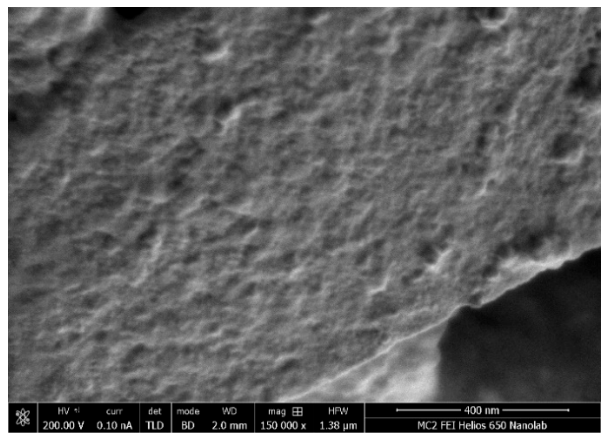

**Figure S2**. SEM images of P@CX particles showing the morphology at various magnifications.

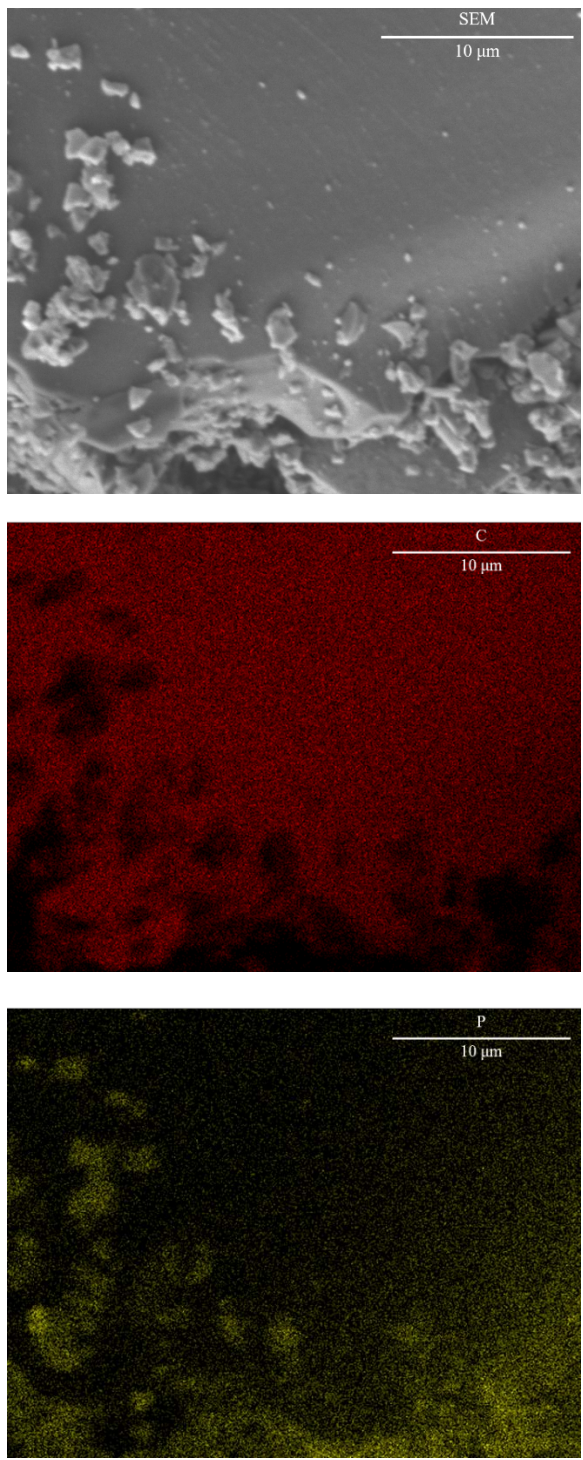

**Figure S3**. EDS mapping of P and C in P@CX particles.

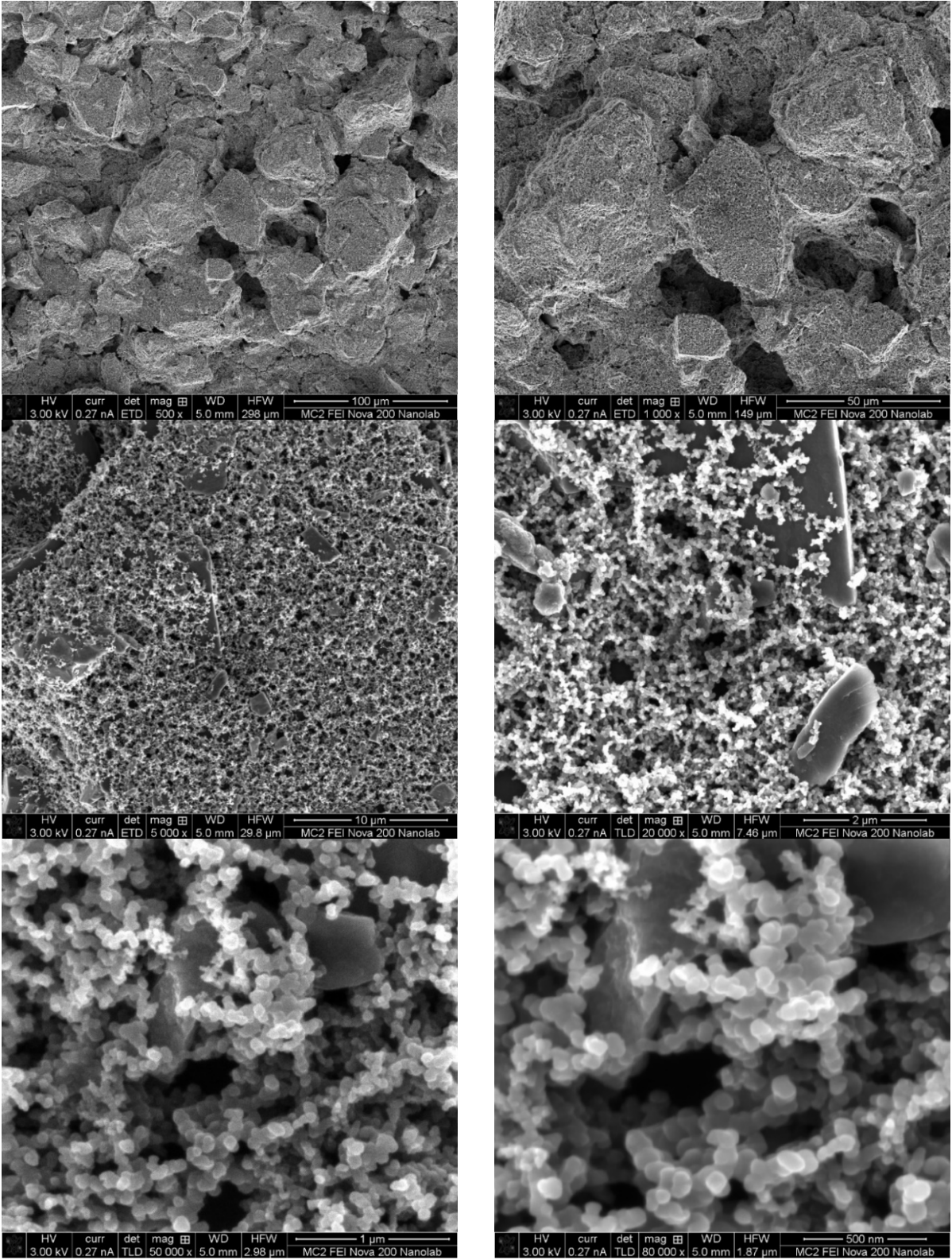

**Figure S4.** SEM images of the P@CX electrode morphology (including carbon black and binder) at various magnifications. The bottom row shows that the P@CX particle surfaces are covered by fine carbon black particles.

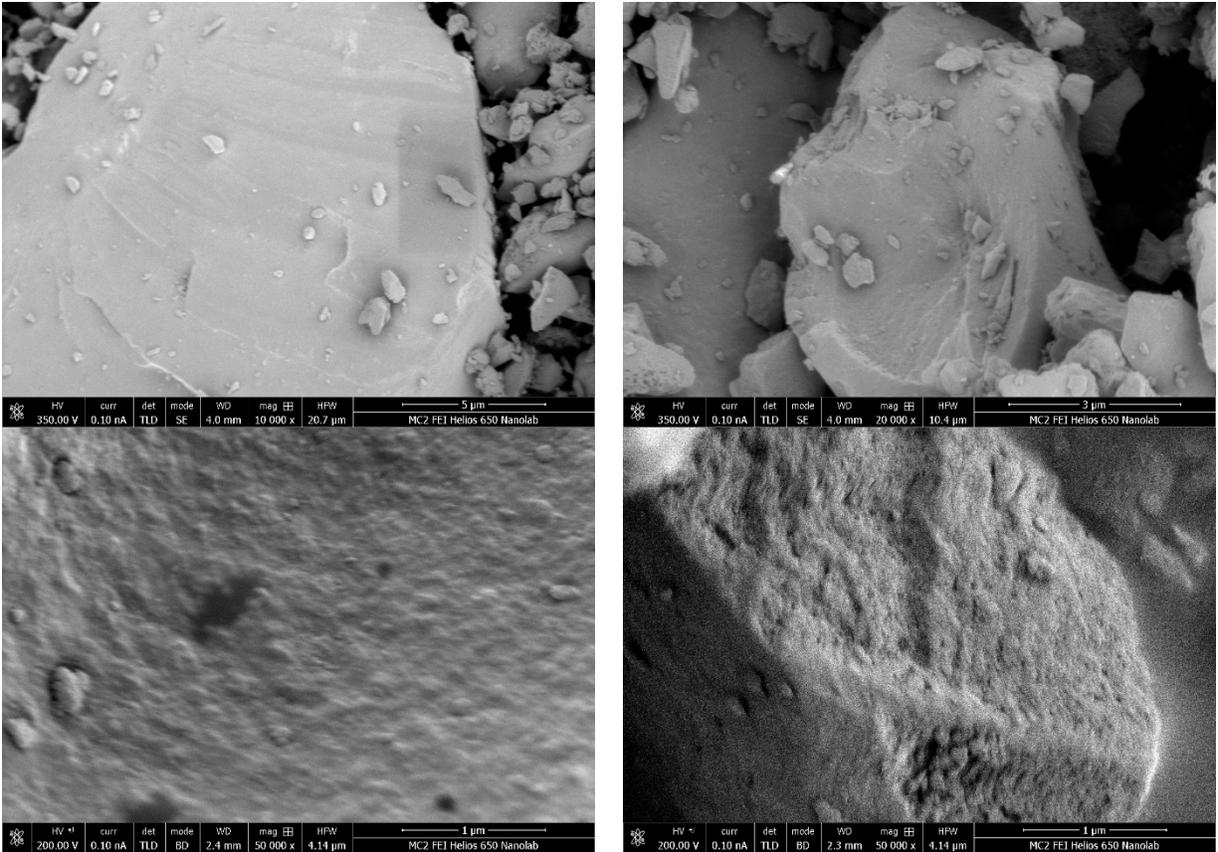

**Figure S5.** SEM images of carbon xerogel (CX) morphology at various magnifications.

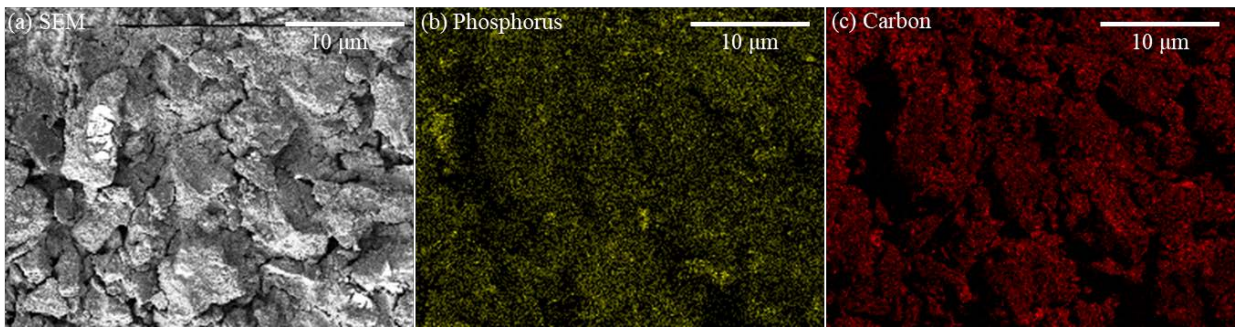

**Figure S6.** EDS mapping of P and C in P@CX electrode after 300 cycles. The capacity is shown in Figure 5d and more SEM images are shown in Figure 6.

6